\documentclass[aps,prl,twocolumn,superscriptaddress,showpacs]{revtex4}
\usepackage{epsf}
\usepackage{amsmath,amssymb}
\usepackage{graphicx}
\usepackage{color}

\begin{document}

\title{How complex is the quantum motion?}


\date{\today}

\affiliation{CNISM, CNR-INFM, and Center for Nonlinear and Complex Systems,
Universit\`a degli Studi dell'Insubria, Via Valleggio 11, 22100
Como, Italy}
\author{Giuliano Benenti}
\affiliation{CNISM, CNR-INFM, and Center for Nonlinear and Complex Systems,
Universit\`a degli Studi dell'Insubria, Via Valleggio 11, 22100
Como, Italy}
\affiliation{Istituto Nazionale di Fisica Nucleare, Sezione di Milano,
Via Celoria 16, 20133 Milano, Italy}
\author{Giulio Casati}
\affiliation{CNISM, CNR-INFM, and Center for Nonlinear and Complex Systems,
Universit\`a degli Studi dell'Insubria, Via Valleggio 11, 22100
Como, Italy}
\affiliation{Istituto Nazionale di Fisica Nucleare, Sezione di Milano,
Via Celoria 16, 20133 Milano, Italy}

\begin{abstract}
In classical mechanics the complexity of a dynamical system is
characterized by the rate of local exponential instability which
effaces the memory of initial conditions and leads to practical
irreversibility. In striking contrast, quantum mechanics appears to
exhibit strong memory of the initial state. Here we introduce a
notion of complexity for a quantum system and relate it to its
stability and reversibility properties.
\end{abstract}


\maketitle

The question of how complex is quantum motion is of fundamental
importance with deep connections to entanglement and decoherence.
However, our knowledge of the relations between complexity, dynamical stability,
reversibility and chaos is far from being satisfactory, sometimes
even confusing and a clearer understanding is necessary.

To this end let us first consider classical motion where things are
quite well settled. Classical complex systems are characterized by
positive Lyapunov exponent i.e. by local exponential instability.
They have positive algorithmic complexity and, in terms of the
symbolic dynamical description, almost all orbits are random and
unpredictable~\cite{ford}.

In spite of many efforts~\cite{prosencomplexity}, the problem of
characterizing the complexity of a quantum system is still open.
Indeed the above notion of complexity cannot be transferred,
\emph{sic et simpliciter}, to quantum mechanics, where there is no
notion of trajectories. Still, a comparison between classical and
quantum dynamics can be made by studying the evolution in time of
the classical and quantum phase space distributions, both ruled by
linear equations.

First investigations focused on \emph{reversibility}, namely on the
propagation of round-off errors in numerical
simulations~\cite{arrow}. Strong and impressive evidence has been
gathered that the quantum evolution is very stable, in sharp
contrast with classical dynamics in which the extreme sensitivity to
initial conditions, which is the very essence of classical chaos,
leads to a rapid loss of memory.

Later on, a different approach focussed on the \emph{stability} properties
of motion under small variations of system parameters. This approach
does not rise any difficulty in the classical context since
exponentially unstable systems exhibit the same rate of exponential
instability by slightly changing the initial conditions with fixed
parameters or by changing parameters with fixed initial conditions.
On the other hand the advantage of the latter approach is that it
can be applied to phase space distributions. Here one computes the
so-called \emph{fidelity}~\cite{peres,prosen}, defined as the
overlap between two distributions evolving under two slightly
different Hamiltonians. It is tempting to connect the behavior of
fidelity to the regular or chaotic behavior of quantum motion.
Indeed the original expectation, which seemed quite natural, was that
fidelity should remain close to one at all times for integrable
systems and fall down exponentially for chaotic
systems~\cite{peresbook,haake}. However, this expectation is not
fulfilled.

For the purpose of the present paper it is necessary here to make
clear the following. The motivation for the introduction of
fidelity, namely the suggestion to analyze the stability of motion
by perturbing the Hamiltonian rather than the initial state,
originated from the observation that in quantum mechanics, due to
the unitary evolution, the scalar product
$\langle\psi(t)|\psi^\prime(t)\rangle$ of two initially close states
$|\psi(0)\rangle$ and $|\psi^\prime(0)\rangle$ does not change in
time. However two points must be stressed: i) the classical
evolution of phase space density is also unitary and linear and
therefore the overlap of two initially close phase space
distributions  does not change with time in classical mechanics as
well; ii) for classically chaotic quantum systems, the fidelity
decay, depending on the perturbation strength, can be Gaussian or
exponential. A power-law decay is also possible in the quantum diffusive
regime~\cite{benenti02}. Furthermore, for
integrable systems the fidelity decay can be faster than for chaotic
systems~\cite{prosen}. On the other hand, even in classical
mechanics the fidelity decay does not clearly distinguish between
chaotic and integrable systems. In short, fidelity  is not a good
quantity to characterize the complexity of motion, neither in
quantum nor in classical mechanics.

In this paper, we propose the \emph{number of harmonics of the
Wigner function} as a suitable measure of the complexity of a
quantum state. We recall that in classical mechanics the number of
harmonics of the classical distribution function in phase space
grows linearly for integrable systems and exponentially for chaotic
systems, with the growth rate related to the rate of local
exponential instability of classical motion~\cite{brumer}. Thus the
growth rate of the number of harmonics is a measure of classical complexity.
Since the phase space approach can be equally used for both
classical and quantum mechanics, the number of harmonics of the
Wigner function appears as the correct quantity to measure the
complexity of a quantum state. In what follows, we examine the
behaviour of this quantity and its relation to fidelity and
reversibility properties. 
Moreover, we show that the number of harmonics can be used to detect
the transition from integrability to quantum chaos.
A detailed derivation of the results
discussed in this paper can be found in Ref.~\cite{harmonicslong}.

{\it The Wigner Function.} -
Let us consider a generic nonlinear system which exhibits a transition
from quasi-integrable to chaotic behavior as the strength of the
nonlinearity is increased. More precisely we consider
the Hamiltonian operator ${\hat H}\equiv H({\hat
a}^{\dag},{\hat a};t)=H^{(0)}({\hat n}={\hat a}^{\dag}{\hat
a})+H^{(1)}({\hat a}^{\dag},{\hat a};t)$ with a time-independent
unperturbed part ${\hat H}^{(0)}$ with a discrete energy spectrum
bounded from below.
Here ${\hat a}^{\dag}, {\hat a}$ are the
bosonic, creation-annihilation operators and $[{\hat a},{\hat
a}^{\dag}]=1\,$.

We will use the method of c-number $\alpha$-phase space borrowed
from quantum optics (see for example~\cite{Glauber63}).
This method is 
basically built upon the basis of the coherent states
$|\alpha\rangle$ which are defined by the eigenvalue problem ${\hat
a}|\alpha\rangle=\frac{\alpha}{\sqrt \hbar}|\alpha\rangle$, where
$\alpha$ is a complex variable independent of the effective Planck's
constant $\hbar$. An arbitrary coherent state is obtained from the
ground state, $|\alpha\rangle={\hat
D}\left(\frac{\alpha}{\sqrt\hbar}\right)|0\rangle$, with the help of
the unitary displacement operator ${\hat
D}\left(\lambda\right)=\exp(\lambda\,{\hat a}^{\dag}-\lambda^*{\hat
a})$.
The Wigner function $W$
in the $\alpha$-phase plane is related to the
density operator $\hat{\rho}$ as follows:
\begin{equation}\label{Wfunc}
W(\alpha^*,\alpha;t)=
\frac{1}{\pi^2\hbar}\int d^2\eta\,
e^{\eta^*\frac{\alpha}{\sqrt\hbar}-
\eta\frac{\alpha^*}{\sqrt\hbar}} {\rm Tr}\left[{\hat\rho(t)}\,{\hat
D(\eta)}\right],
\end{equation}
where the integration runs over the complex $\eta$-plane. 

The harmonic's amplitudes $W_m$ of the Wigner function are
given by the expansion
\begin{equation}\label{Four}
W(\alpha^*,\alpha;t)=\frac{1}{\pi}\,\sum_{m=-\infty}^{\infty}W_m(I;t)\,
e^{i m\theta}\,,
\end{equation}
where $\alpha=\sqrt{I}\,e^{-i\theta}\,$, 
with $(I,\theta)$ action-angle variables. 

{\it The fidelity.} -
Following the approach developed in Ref.~\cite{ikeda}, we consider
now the forward evolution
\begin{equation}\label{Probe}
{\hat\rho}(t)={\hat U(t)}{\hat\rho}(0){\hat U}^\dagger(t)
\end{equation}
of an initial (generally mixed) state ${\hat\rho}(0)$ up to some
time $t=T$. A perturbation ${\hat P}(\xi)$ is then applied at this
time, with perturbation strength $\xi$. For our purposes, it will be
sufficient to consider unitary perturbations ${\hat P}(\xi)=e^{-i\xi
{\hat V}}$, where ${\hat V}$ is a Hermitian operator. The perturbed
state
\begin{equation}
{\hat{\tilde\rho}}(T,\xi)={\hat P}(\xi){\hat\rho}(T){\hat
P}^\dagger(\xi)
\end{equation}
is then evolved backward, with the same Hamiltonian, 
for the time $T$, thus obtaining the final
state
\begin{equation}
{\hat{\tilde\rho}}(0|T,\xi)={\hat
U}^{\dag}(T){\hat{\tilde\rho}}(T,\xi){\hat U}(T).
\end{equation}

Finally, we consider the distance between the reversed
${\hat{\tilde\rho}}(0|T,\xi)$ and the initial ${\hat\rho}(0)$ state,
as measured by the Peres fidelity~\cite{peres}
\begin{equation}\label{PFid}
F(\xi;T)=\frac{{\rm
Tr}[{\hat{\tilde\rho}}(0|T,\xi){\hat\rho(0)}]} {{\rm
Tr}[{\hat\rho}^2(0)]}
=\frac{{\rm Tr}[{\hat{\tilde\rho}}(T,\xi){\hat\rho(T)}]} {{\rm
Tr}[{\hat\rho}^2(T)]}.
\end{equation}
This quantity is bounded in the interval
$[0,1]$ and
the distance between the initial and the time-reversed state
is small when $F(\xi;T)$ is close to one.
In particular, $F(\xi;T)=1$ when the two states coincide.
The last equality in (\ref{PFid}) is a consequence of the unitary time
evolution.

The Peres fidelity (\ref{PFid}) can be expressed in terms of the
Wigner function as
\begin{equation}\label{FidW}
\begin{array}{c}
{\displaystyle
F(\xi;T)=\frac{\int d^2\alpha\,W\left(\alpha^*,\alpha;0\right)
\tilde{W}\left(\alpha^*,\alpha;0\big|T,\xi\right)}
{\int d^2\alpha\,W^2\left(\alpha^*,\alpha;0\right)}=
}
\\
{\displaystyle
\frac{\int d^2\alpha\,W\left(\alpha^*,\alpha;T\right)
\tilde{W}\left(\alpha^*,\alpha;T,\xi\right)} {\int
d^2\alpha\,W^2\left(\alpha^*,\alpha;T\right)}\,.
}
\end{array}
\end{equation}
The advantage of this representation is
that it remains valid in the classical case when the Wigner
function reduces to the classical phase-space distribution function
$W_c(\alpha^*,\alpha;t)\,$.

{\it Growth of the number of harmonics.}-
As is well known, the paramount property of the classical
dynamical chaos is the exponentially fast structuring of the
system's phase space on finer and finer scales. In particular, the
number ${\cal M}(t)$ of harmonics (Fourier components)
that significantly contribute in
the expansion (\ref{Four}) of the classical phase space
distribution $W_c$, grows exponentially in time for chaotic motion.
The crucial point is that only in quantum
mechanics the number of harmonics of the Wigner function is directly
related to the expectation value of  physical observables. Therefore in
quantum mechanics an exponential growth of number of harmonics is
not allowed in general~\cite{chirikov,gu,brumer}.

For an explicit numerical evaluation of the number of harmonics
we will focus on the quantity
\begin{equation}\label{Aver_m}
\langle m^2\rangle_t = \frac{\sum_{m=-\infty}^{+\infty}m^2
\int_0^\infty dI |W_m(I;t)|^2}{\sum_{m=-\infty}^{+\infty}
\int_0^\infty dI |W_m(I;t)|^2}.
\end{equation}
The quantity $\sqrt{\langle m^2\rangle}_t$
provides an estimate of the
number ${\cal M}(t)$ of harmonics developed by the time $t$ and therefore
of the complexity of the Wigner function at time $t$.

{\it Relation between fidelity and number
of harmonics.} -
This relation takes
a very simple form when the perturbation at the reversal time $t=T$
is a rotation of the quantum phase-space distribution
$W(\alpha^*,\alpha;t)$. Consider the unitary transformation
(\ref{Probe}) with the perturbation operator ${\hat V}={\hat n}$ and
the rotation angle $\xi$. In this case we obtain~\cite{harmonicslong}
\begin{equation}\label{Fid_t}
F(\xi;t)=1-2\frac{\sum_{m=-\infty}^{+\infty}\sin^2\left(\xi m/2\right)
\int_0^\infty dI |W_m(I;t)|^2}{\sum_{m=-\infty}^{+\infty}
\int_0^\infty dI |W_m(I;t)|^2}.
\end{equation}
The lowest-order $\xi$-expansion of this equation reads
\begin{equation}\label{Fid_tlinear}
F(\xi;t)\approx 1-\frac{1}{2}\xi^2\,\langle m^2\rangle_t.
\end{equation}

Notice that relations (\ref{Fid_t}) and (\ref{Fid_tlinear}) between
distance $F$ and number of harmonics can be applied to classical
dynamics, provided that the harmonics of the classical distribution
function $W_c$ instead of those of the Wigner function $W$ are used.

{\it Illustrative example-}. 
Let us consider the kicked quartic
oscillator model, defined by the
Hamiltonian~\cite{Berman78,chirikov,Sokolov84,Sokolov07}
\begin{equation}\label{Ham}
{\hat H}=\hbar\,\omega_0 {\hat n}+\hbar^2\,{\hat
n}^2-\sqrt{\hbar}\,g(t)({\hat a}+{\hat a}^{\dag}),
\end{equation}
where $g(t)=g_0\sum_s\delta(t-s)$,
${\hat n}={\hat a}^{\dag}{\hat
a}$, $[{\hat a},{\hat a}^{\dag}]=1$. In our units, the time and
parameters $\hbar, \omega_0$ as well as the strength $g_0$ of the driving
force are dimensionless. The period of the driving force is set to
one. The corresponding classical Hamiltonian function,
expressed in terms of complex canonical variables $\alpha,
i{\alpha}^*$, is given by
\begin{equation}
\label{Hamc}
H_c=\omega_0|\alpha|^2+|\alpha|^4-g(t)(\alpha^*+ \alpha)\,.
\end{equation}

We choose the initial state
to be  an isotropic mixture of coherent states:
${\hat\rho}(0)= \int d^2 \alpha\, {\cal
P}(|\alpha|^2)|\alpha\rangle
\langle\alpha|$,
where
${\cal P}(I=|\alpha|^2)
=\frac{1}{\pi\Delta}\,e^{-I/\Delta}$.
Correspondingly, the initial Wigner function is isotropic and Gaussian,
$W(\alpha^*,\alpha;0)\propto 
e^{-\frac{|\alpha|^2}{\Delta+\hbar/2}}$.
Only the $m=0$ harmonic is excited, so that $\langle m^2 \rangle_{t=0}=0$.
The particular case $\Delta=0$ corresponds to the pure ground state, which
occupies the minimal quantum cell with the area $\hbar/2\,.$

The classical dynamics of model (\ref{Ham}) becomes chaotic 
(with negligible stability islands) when the
perturbation strength $g_0\gtrsim1$.
In this regime the mean action grows diffusively with the
diffusion coefficient $D\approx g_0^2.$

\begin{figure}[!t]
\includegraphics[width=8.0cm,angle=0]{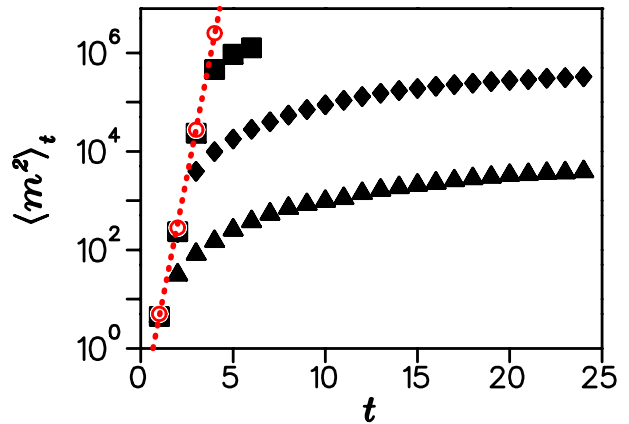}
\caption{(color online) Root-mean-square radius $\langle m^2 \rangle_t$
of the distribution of harmonics versus time $t$,
at $\omega_0=1$, $g_0=1.5$, $\delta=0.5$. Squares, diamonds
and triangles correspond to $\hbar = 0.01, 0.1$ and 1.
In this latter case, $\hat{\rho}(0)=|0\rangle\langle 0|$.
Empty circles refer to classical dynamics and the dashed
line fits these data.}
\label{fig:Wquantclass}
\end{figure}

{\it Chaotic regime.} -
We now compare, in the chaotic regime, 
the evolution in time of $\langle m^2 \rangle_t$ for quantum
and classical dynamics. For this purpose, we
solve both the quantum and the classical
Liouville equation. In the latter case, the initial phase
space distribution $W_c(\alpha^*,\alpha;0)\propto 
e^{-\frac{|\alpha|^2}{\delta}}$ has size
$\delta$ which coincides,
for a given value of $\hbar$, with the size
$\hbar/2$ of the Wigner function
corresponding to the initial quantum ground state
${\hat{\rho}}(0)=|0\rangle\langle 0|$.
The quantum to classical transition is then explored
by keeping $\delta$ constant and
considering, for smaller and smaller values of $\hbar$,
initial incoherent
mixtures of size $\delta=\Delta+\hbar/2$.
The results are shown in Fig.~\ref{fig:Wquantclass}.
The exponential increase of $\langle m^2 \rangle_t$
takes place only up to the Ehrenfest time scale
$t_E \propto \ln \hbar$~\cite{Berman78},
consistently with the findings reported in Refs.~\cite{gu,brumer}.
After that time, a much slower power-law increase
follows. Namely the number of harmonics
${\cal M}(t)\sim \sqrt{\langle m^2 \rangle}$
increases linearly for the pure state case,
where $\langle m^2 \rangle_t \approx \langle n \rangle_t^2
\propto t^2$ due to diffusive growth of the mean action,
and slower than linearly for mixtures~\cite{harmonicslong}.
This growth eventually saturates
due to quantum localization~\cite{chirikov} of diffusive motion.

\begin{figure}[!t]
\includegraphics[width=8.0cm,angle=0]{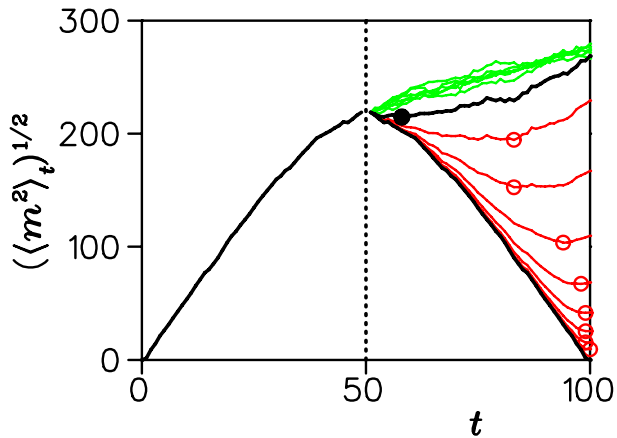}
\caption{(color online) Reversibility properties of quantum dynamics.
The backward evolution starts at the reversal time $T=50$.
We show $\sqrt{\langle m^2 \rangle_t}$, for different
values of the perturbation parameter: from bottom to top,
$\xi={\xi}_c(T)\times \exp{(-l/2)}$,
$l=8,\ldots,1$, $l=0$
(thick black curve marked by the closed circle), and $l=-1,\ldots,-6$,
at $\omega_0=1$, $\hbar=1$, $g_0=2$, $\Delta=1$.
Circles indicate positions of the minimum on each curve.}
\label{fig:reversibility}
\end{figure}

From Eq.~(\ref{Fid_tlinear}) we can estimate
a critical perturbation strength
$\xi_c(T)\approx \sqrt{2/\langle m^2\rangle_T}$,
such that the fidelity $F(\xi;T)$ remains
close to 1 after the backward evolution as long as
$\xi\ll \xi_c(T)$, whereas reversibility is lost when
$\xi\gtrsim\xi_c(T)$.
This statement is illustrated
in Fig.~\ref{fig:reversibility}
for the number of harmonics,
estimated by $\sqrt{\langle m^2 \rangle_t}$.
Therefore, we establish a direct connection between 
complexity of phase space distribution and degree of 
reversibility of motion.
Due to the strikingly different growth in time of the number of
harmonics for classical and quantum chaotic motion, $\xi_c(T)$ drops
exponentially with $T$ in the classical case and at most linearly in
the quantum case (after the Ehrenfest time scale). Therefore, 
our analysis explains the 
numerically observed~\cite{arrow} much weaker sensitivity of quantum
dynamics to perturbations as compared to classical dynamics.

Note that, due to the exponential proliferation of the number of
harmonics in classical mechanics, formula (\ref{Fid_tlinear})
is only valid up to a time logarithmically short in
the perturbation strength $\xi$. After that time, due to
diffusive growth of the mean action, the decay of the fidelity
$F$ turns from exponential to power law~\cite{benenti03},
$F \propto 1/(Dt)$, while the number of harmonics still grows
exponentially with time and correctly describes the complexity
of chaotic motion.

\begin{figure}[!t]
\includegraphics[width=8.0cm,angle=0]{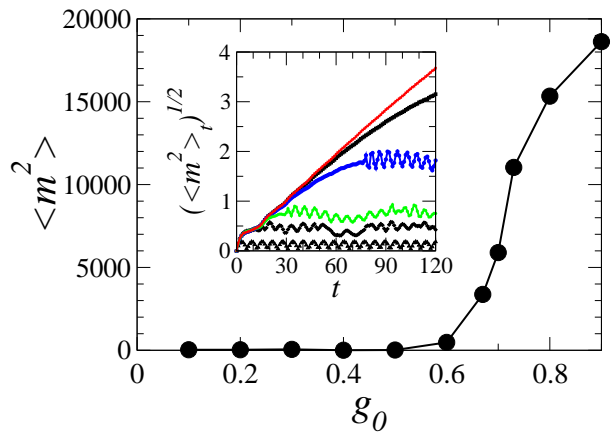}
\caption{(color online) Dependence of $\langle m^2 \rangle_t$
on the perturbation strength $g_0$ at time $t=3$, for $\hbar=0.01$,
$\omega_0=1$, $\Delta=0$. 
Inset: time evolution of $\sqrt{\langle m^2 \rangle_t}$
in the integrable regime, at $g_0=1$, $\omega_0=1$, 
$\delta=0.5$ and, from bottom to top,
$\hbar=1,0.1,0.05,0.02,0.01,0.005$.} 
\label{fig:transition}
\end{figure}

{\it Crossover from integrability to quantum chaos.} -
It is known that in the integrable regime the number of harmonics,
computed in the action-angle representation, grows linearly 
with time~\cite{brumer} and that nearby orbits separate linearly
fast~\cite{CCF}. We show (see the inset of Fig.~\ref{fig:transition})
that the linear growth of $\sqrt{\langle m^2 \rangle}$ also takes 
place in quantum mechanics, up to the Heisenberg time scale 
$t_H \propto \hbar^{-1}$.   
The strikingly different behavior of the number of harmonics
of the Wigner function in the integrable and chaotic regime 
suggests that this quantity can be used 
to detect the transition to quantum chaos.
We show in Fig.~\ref{fig:transition} 
that $\langle m^2 \rangle$, computed at a given time $t$, 
exhibits a sharp increase when the perturbation parameter 
$g_0\gtrsim 0.5$. From this figure we can conclude that 
the crossover from integrability to chaos takes place 
in the region $0.5 \lesssim g_0 \lesssim 0.7$.

{\it Discussion.} -
To summarize, we have shown that the number of harmonics of
the Wigner function is a suitable measure of complexity of
a quantum state, in that this quantity is directly related to the
reversibility properties of quantum motion and, at the classical
limit, reproduces the well-known notion of complexity based
on local exponential instability.
We would like to stress that in relation to other, very
interesting, proposed measures 
of quantum complexity, such as quantum dynamical 
entropies~\cite{prosencomplexity}, our quantity is very
convenient for numerical investigations. 
It becomes therefore possible to investigate complexity as a
function of the effective Planck's constant.
Moreover, the above
outlined phase-space approach is quite general and can be
readily extended to systems with
arbitrary number of degrees of freedom, including qubit systems,
whose Hamiltonian can be
expressed in terms of a set of bosonic creation-annihilation
operators. Therefore, in many-body systems 
the (number of) harmonics of the Wigner function could shed some
light on the connection between complexity and entanglement,
a fundamental issue of great relevance for the prospects
of quantum information science. Moreover, we believe that the dependence of
the number of harmonics on control parameters could be used not only
to investigate the integrability to quantum chaos crossover
but also to detect quantum phase transitions.


\vspace{-0.6cm}

\begin{thebibliography}{01}

\bibitem{ford}
J. Ford, Phys. Today, April 1983, pag. 40;
%
V.M. Alekseev and M.V. Jacobson,
Phys. Rep. \textbf{75}, 287 (1982).

\bibitem{prosencomplexity}
R. Alicki and M. Fannes, {\it Quantum dynamical systems} (Oxford
University Press, 2001); T. Prosen, J. Phys. A \textbf{40}, 7881
(2007), and references therein.

\bibitem{arrow}
D.L. Shepelyansky, Physica D {\bf 8}, 208 (1983); 
G. Casati {\it et al.},
Phys. Rev. Lett. {\bf 56}, 2437 (1986).

\bibitem{peres}
A. Peres, Phys. Rev. A {\bf 30}, 1610 (1984).

\bibitem{prosen}
T. Gorini {\it et al.},
Phys. Rep. {\bf 435}, 33 (2006).

\bibitem{peresbook}
A. Peres,
{\it Quantum theory: Concepts and Methods}
(Kluwer Academic, Dordrecht, 1993).

\bibitem{haake}
F. Haake,
{\it Quantum signatures of chaos} (2nd. Ed.)
(Springer-Verlag, 2000).

\bibitem{benenti02}
G. Benenti and G. Casati, Phys. Rev. E {\bf 65}, 066205 (2002);

\bibitem{brumer}
A.K. Pattanayak and P. Brumer, Phys. Rev. E {\bf 56}, 5174 (1997);
J. Gong and P. Brumer, Phys. Rev. A {\bf 68}, 062103 (2003).

\bibitem{harmonicslong}
V.V. Sokolov {\it et al.},
Phys. Rev. E {\bf 78}, 046212 (2008).

\bibitem{Glauber63}
R.J. Glauber, Phys. Rev. {\bf 131} 2766 (1963);
%
G.S. Agarwal and E. Wolf, Phys. Rev. D
{\bf 2}, 2161 (1970); {\it ibid.} 2187.

\bibitem{ikeda}
K.S. Ikeda, in {\it Quantum chaos: between order and disorder}, G.
Casati and B.V. Chirikov (Eds.) (Cambridge University Press, 1995).

\bibitem{chirikov}
B.V. Chirikov, F.M. Izrailev, and D.L. Shepelyansky, Sov. Sci. Rev.
C {\bf 2}, 209 (1981).

\bibitem{gu}
Y. Gu, Phys. Lett. A {\bf 149}, 95 (1990).

\bibitem{Berman78} G.P. Berman, G.M. Zaslavsky, Physica A
{\bf 91}, 450 (1978); {\it ibid.} {\bf 97}, 367 (1979).

\bibitem{Sokolov84} 
V.V. Sokolov, 
Sov. J. Theor. Math. {\bf 61}, 104 (1985).

\bibitem{Sokolov07} V.V. Sokolov, G. Benenti,
and G. Casati, Phys. Rev. E {\bf 75}, 026213 (2007).

\bibitem{benenti03}
G. Benenti, G. Casati, and G. Veble, Phys. Rev. E
{\bf 67}, 055202(R) (2003).

\bibitem{CCF}
G. Casati, B.V. Chirikov, and J. Ford, 
Phys. Lett. {\bf 77A}, 91 (1980).

\end{thebibliography}
\end{document}